\documentclass{article}

\usepackage{arxiv}

\usepackage[utf8]{inputenc} % allow utf-8 input
\usepackage[T1]{fontenc}    % use 8-bit T1 fonts
\usepackage{hyperref}       % hyperlinks
\usepackage{url}            % simple URL typesetting
\usepackage{booktabs}       % professional-quality tables
\usepackage{amsfonts}       % blackboard math symbols
\usepackage{nicefrac}       % compact symbols for 1/2, etc.
\usepackage{microtype}      % microtypography
\usepackage{lipsum}		% Can be removed after putting your text content
\usepackage{graphicx}
\usepackage[square,sort,comma,numbers]{natbib}
\usepackage{doi}
\usepackage{tablefootnote}
\usepackage{authblk}

\title{\LARGE \bf
	Mining and evaluation of patients’ diagnostic-therapeutic paths through state sequences analysis
}
\author{{ Laura Savaré\textsuperscript{1,}\textsuperscript{2,}\textsuperscript{3}, Francesca Ieva\textsuperscript{1,}\textsuperscript{2,}\textsuperscript{3}, Giovanni Corrao\textsuperscript{3}, Antonio Lora\textsuperscript{3,}\textsuperscript{4,}}\\
{\textsuperscript{1} MOX - Department of Mathematics, Politecnico di Milano, Milan, Italy,\\
\textsuperscript{2} CHDS,  Health Data Science Center, Human Technopole, Milan, Italy\\
\textsuperscript{3} National Centre for Healthcare Research \& Pharmacoepidemiology, University of Milano- Bicocca, Milan, Italy\\
\textsuperscript{4} Department of Mental Health and Addiction Services, ASST Lecco, Lecco, Italy\\
}}

\renewcommand{\shorttitle}{State sequences analysis}

\begin{document}
\maketitle
\begin{abstract}
The concept of care pathways is increasingly being used to enhance the quality of care and to optimize the use of resources for health care. Nevertheless, recommendations regarding the sequence of care are mostly based on consensus-based decisions as there is a lack of evidence on effective treatment sequences. In a real-world setting, classical statistical tools resulted to be insufficient to adequately consider a phenomenon with such high variability and has to be integrated with novel data mining techniques suitable of identifying patterns in complex data structures. Data-driven techniques can potentially support the empirical identification of effective care sequences by extracting them from data collected routinely. The purpose of this study is to  perform sequence analysis to identify different patterns of treatment and to asses the most efficient in preventing adverse events. 
The clinical application that motivated the study of this method concerns the several problems frequently encountered in the quality of care provided in the mental health field. In particular, we analyzed administrative data provided by Regione Lombardia related to all the beneficiaries of the National Health Service with a diagnosis of schizophrenia from 2015 to 2018 resident in Lombardy, a region of northern Italy. In fact, the care pathways of patients affected by severe mental disorders often do not correspond to the standards required by the guidelines in this field. 
This methodology considers the patient’s therapeutic path as a conceptual unit, i.e., a sequence, composed of a succession of different states that can describe longitudinal patient's status. In this work, we define the states to be the weekly coverage of different treatments (e.g., psychiatric visits, psychosocial interventions, and anti-psychotic drugs), and we use the longest common subsequences (dis)similarity measure to compare and cluster the sequences. This kind of information, such as common patterns of care that allowed us to risk profile patients, can provide health policymakers an opportunity to plan optimum and individualized patient care by allocating appropriate resources, analyzing trends in the health status of a population, and finding the risk factors that can be leveraged to prevent the decline of mental health status at the population level.
\end{abstract}

% keywords can be removed

\newpage
\section{Introduction}
Diagnostic-therapeutic pathways are evidence-based interventions aimed to organize the assistance process for specific groups of patients, and to enhance the quality of care across the continuum by improving risk-adjusted patients outcomes, promoting patient safety, increasing patient satisfaction, and optimizing the use of resources \cite{0}. Recommendations regarding the sequence of care are mostly based on consensus-based decisions as there is a lack of evidence on effective treatment sequences and a consequent high variability in treatments. In chronic diseases, studying the effect of these patterns on adverse outcomes is of clinical relevance. Nevertheless, the way they are accounted for into predictive models is far from being informative as it may be \cite{1,2,3}.
In the field of epidemiological research, the use of administrative health databases has now become a widely used strategy thanks to the always greater reliability
of the detection methodologies adopted, which involve the acquisition of high quality data \cite{A}. They also offer an ever increasing amount of useful information that allow  to conduct epidemiological studies with
less resources and cost and time savings. One of the main features of the studies conducted with such databases concerns the possibility of reflecting real clinical practice on large and unselected populations, an aspect that surely exceeds one of the main limits of randomized clinical trials. Although they offer a growing amount of useful information, to date, studies focusing on individual pathways have mainly remained descriptive, without taking into account the possible evolution of care consumption over time \cite{B,C}. Although in recent years some methodologies were proposed to tackle this kind of information (such as the trajectories models \cite{D}, temporal association rules \cite{E}, supervised machine learning \cite{F}, the tree-based scan statistic \cite{G}, and the latent class model \cite{H,I,L}), they are not always suitable to evaluate complex longitudinal patterns of care and to be implemented on real-world data. For example, they usually work on a limited number of class-defining variables and are thus not suitable to evaluate complex longitudinal patterns of prescriptions. Furthermore, pragmatic aspects such as the need for stakeholders to understand the method in order to be confident in the results must be considered. 
In this framework, classical statistical tools have to be integrated with novel data mining techniques for properly identifying patterns in complex data structures. Data-driven techniques can potentially support the empirical identification of effective care sequences by extracting them from data collected routinely in health care \cite{M}.

\hspace{20pt} State Sequence Analysis (SSA), is an upcoming technique in epidemiology, derived from social sciences, could provide useful insights on the chronology of care consumption, the interval and timing of treatment patterns used in the clinical practice and the effectiveness of different treatment sequences. SSA also allows for identification of specific patterns \cite{Mbis}. Specifically, attention is focused on the ordered sequence of states (or activities) experienced by individuals over a given time span (usually at T equally spaced discrete time periods). To this end, pairwise dissimilarities among sequences in their entirety are first assessed. Dissimilarity matrices are then employed to identify the most typical trajectories using cluster analysis. Applied to data recorded in health care information systems, field of application which is still almost unexplored, this technique can assess performed medical behavior and the chronological sequence of these behaviors, allowing to identify the most relevant patterns in the data and to asses the most efficient in preventing adverse events. 

\hspace{20pt} In this work, a novel pipeline for using SSA in the health area in order to monitor and evaluate the diagnostic-therapeutic paths of patients is proposed and discussed. It is then applied to the pathway of patients affected by schizophrenic disorder. The choice of this application is motivated by the fact that, in mental health, problems are frequently encountered in the quality of care provided \cite{N,O}. The care pathways of patients with severe mental disorders often do not correspond to the standards required by the evidence in this field, and they present a high variability between countries and within them. Moreover, the lack of definition of optimal diagnostic-therapeutic paths has led us to investigate if a common pattern of care can be recognised and if a risk profiling of the patients can provide health policy makers an opportunity to plan optimum and individualised patient care by allocating appropriate resources, analysing trends in the health status of a population, and finding the risk factors that can be modified to prevent decline in the health status at a population level. In a real-world setting, state of the art tools resulted to be insufficient to adequately consider a phenomenon with such high variability \cite{Obis, Obisbis}, and it is also difficult to find the statistical methodology suitable to properly consider so many different items. For these reasons, it is very important to outline synthetic tools in order to provide process indicators.

In particular, there is the need to describe and analyse longitudinal care pathways to then evaluate their association with the incidence of negative outcomes. Indeed, only in recent years European countries increasingly recognised how important it is to equip policy makers, care providers and service users in health system with tools to move towards a person-centered system, and improvement is being made on what it actually means to deliver person-centered mental health care. This became a priority also at European level, as testified by a number of project and initiatives aimed at pursuing these goals, such as the JA ImpleMENTAL, an European Joint Action born with the purpose of providing Support for Member States’ implementation of best practices in the area of mental health. In fact, in Europe, only a few countries currently use indicators to routinely assess the quality of mental health care and real-world data collected in HCU databases may represent essential leverage for quality improvement of mental health care. Right now, there are very few attempts to build these indicators to monitor services and an evaluation process is not yet used in the routine practice \cite{P}.
Therefore, clinical indicators are urgently needed since they are a useful tool to address the purpose of monitoring and evaluating the quality of care provided by NHS, specifically in the field of mental health \cite{Pbis, Pbisbis}. 
The aim of the present work is to provide a reference method to allow the governance to monitor over time, assess the quality and optimize the diagnostic-therapeutic pathways. This knowledge could be used for (re)designing and optimizing existent care pathways.

\hspace{20pt} The paper is structured as follows. Section 2 introduces the SSA method, and its steps: the data coding and the pathway construction (Section 2.1),  the representation of the sequences (Section 2.2), the description of the unidimensional indicators summarizing longitudinal characteristics of the sequences (Section 2.3), the dissimilarity measures to compare sequences and the sequences profiling via unsupervised procedures (Section 2.4), and finally the post-hoc analysis and prediction models using sequences' clusters and unidimensional indicators (Section 2.5). Data extraction, inclusion criteria, and the study setting are described in Section 3.  Key results from applying these methods to administrative data for patients affected by schizophrenic disorder are presented in Section 4. In Section 5, we end with a discussion of the strengths of representing diagnostic-therapeutic pathways as sequences through administrative data in a state sequence analysis framework and limitations of the current approach, which open doors to further developments in this area.

\hspace{20pt} Statistical analyses were performed in the R-software environment \cite{Pbis3}, using \texttt{TraMineR} package for mining and visualizing sequences of categorical data \cite{Pbis4}. R codes are publicly available at \texttt{https://github.com/lsavare/SSA}.

\section{Methods: The SSA Approach for assessing patterns of care}
This work is concerned with categorical sequence data and more specifically with state sequences, where the position of each successive state receives a meaningful interpretation in terms of age, date, or more generally of elapsed time or distance from the beginning of the sequence. Using SSA terminology, a care pathway corresponds to a sequence during which successive “states”, i.e., the value taken by the variable at a time t, are encountered. The term “sequence” will hence be used to refer to the entire care pathway, and applied to data recorded in health care information systems, can assess performed diagnostic-therapeutic path of patients. The primary objective of sequence methods is then to extract simplified workable information from sequential data sets; that is, to efficiently summarize and render these sets and to categorize the sequential patterns into a limited number of groups. This is essentially an exploratory task that consists of computing summary indicators, as well as sorting, grouping and comparing sequences. The resulting groups and real-value indicators may then be submitted to classical inferential methods and serve, for instance, as response variables or explanatory factors for regression-like models.\\

\subsection{Object oriented representation: identifying of states and sequences}

\begin{figure}[!h]
    \centering
    \includegraphics[scale=0.60]{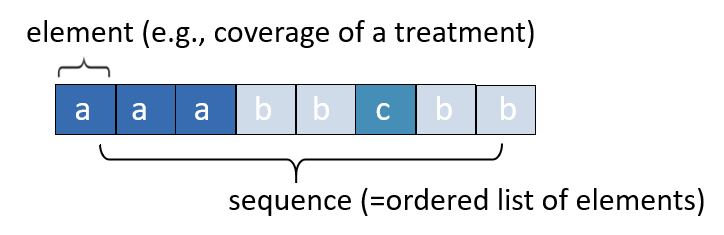}
    \caption{Example of sequence}
    \label{fig:1}
\end{figure}

State sequences can be represented in many different ways, depending on the data source and on how the information is organized \cite{Pbis5, Pbis6}. Usually, to carry out SSA, it is necessary first to convert data into a state sequence format, where each row represents a conceptual unit (e.g., a patients). Data organization and conversion between formats are discussed in detail in \cite{R}, where an ontology of longitudinal data presentations is given.
Formally, a state sequence of length \textit{l} is defined as an ordered list of \textit{l} elements successively chosen from finite set A of size $a = \left|{A}\right|$, namely \textit{alphabet}, that includes all the possible states appearing into the data. An element can be a certain status (e.g., employment, health status or coverage of a treatment), a physical object (e.g. base pair of DNA, protein, or enzyme), or an event (e.g. an hospitalization). The positions of the elements are fixed, ordered by elapsed time and they refer to a relative, not absolute, time point. A natural way of representing a sequence x is by listing the successive elements that form the sequence $x = (x_1, x_2, . . . , x_\textit{l})$, with $x_j \in  A$. (Figure \ref{fig:1}). The addressed methods are for sets of sequences each of which is considered as a whole; i.e., as a conceptual unit. 
Once the alphabet has been specified, it is necessary to define the length of the study period and the time unit (e.g., day, week, month, etc.) of the analysis, also called \textit{sequence granularity}. These two aspects will determine the start, the end, and, therefore, the length of the sequences to be analyzed. Including the sequential information in the research design increases the complexity of the analysis because the number of possible sequences grows exponentially with the sequence length. Dealing with sequence data therefore raises two questions: (i) how can the sequential character of the data be maintained without reducing it to single events, (ii) and how can the variation in the sequences be optimally reduced. SSA can answer these questions through  the visualization of the sequences, the calculation of some indicators for the characteristics of each sequence, the comparison between different sequences using distance measures and grouping of them. 

\subsection{Defining process indicators via suitable sequence representation}
Sequences may be visualized through a sequence index plot. It represents them by horizontally stacked boxes that are colored according to the state at the successive positions. We thus visualize, for each case, the individual longitudinal succession of states as well as, through the length of each color segment, the duration spent in each state. The alignment also permits easy transversal comparisons at each position. However, when it is used on a large number of sequences, it fails to capture specific patterns and it is also difficult to interpret. Others options to visualize sequences are the frequency and representative plots \cite{Pbis4}. In sequence frequency plots, sequences are sorted according to their frequency in the data set and usually the 10 most frequent sequences with bar width proportional to the frequencies are plotted, while in representative sequence plots, sequences are sorted according to their representativeness score. Finally, the state distribution plot shows the proportion, at each time, of the states. An interesting summary that can be derived from the state distributions is the sequence made of the most frequent state at each position (i.e. modal states) \cite{Pbis4}.

\subsection{Sequences' longitudinal characteristics}
Other important features of the sequences can be described by the number of states in each sequence, the mean time spent in each state, and the number of transitions between states. Moreover, some indicators are also defined to summarize longitudinal characteristics of individual sequences, such as entropy, turbulence and complexity of the sequence \cite{R}. In particular, we used the longitudinal Shannon entropy that can be viewed as a measure of uncertainty in predicting the next state within a sequence of data:
\begin{equation}
    h(\pi_1,...,\pi_\alpha)= -\sum_{i=1}^{a}\pi_i log(\pi_i)
\end{equation}
where \textit{a} is the size of the alphabet and $\pi_i$ the proportion of occurrences of the \textit{i}th state in the considered sequence. Therefore, entropy is 0 when all the states of a sequence are the same; it is maximum when a sequence is composed of all states in equal proportion.\\
Sequence turbulence is a measure proposed by Elzinga \& Liefbroer (2007). It is based on the number $\phi(x)$ of distinct subsequences that can be extracted from the distinct successive state sequence and the variance of the consecutive times $t_i$ spent in the distinct states. For a sequence x, the formula is
\begin{equation}
     T(x) = log_2 \bigg{(}\phi(x)\frac{s^2_{t,max}(x)+1}{s^2_t(x)+1}\bigg{)}
\end{equation}
where $s_t^2(x)$ is the variance of the successive state durations in sequence $x$ and $s_{t,max}^2(x)$ is the maximum value that this variance can take given the number of states and the total duration of the sequence. From a prediction point of view, the higher the differences in state durations and hence the higher their variance, the less uncertain the sequence. In that sense, small duration variance indicates high complexity.

\subsection{Measuring dissimilarity and clustering sequences}
Dissimilarity measures among sequences can be classified into measures based on the (minimal) cost of transforming one sequence into the other and those defined as the count of matching attributes.
Another interesting distinction is between those that make position-wise comparisons; i.e., that do not allow shifting a sequence or part of it, and those accounting for similar shifted patterns \cite{S}. 
The optimal matching (OM) algorithm is the most widely known technique in the social sciences \cite{Sbis}. OM is a family of dissimilarity measures derived from what was originally proposed in the field of information theory and computer science by Levenshtein (1965) and adapted to life course analysis by Abbott (1995) \cite{Sbis2}. Basically, OM expresses distances between sequences in terms of the minimal amount of effort, measured in terms of subsequent operations, that is required to transform one sequence into the other one. Three basic operations to transform sequences are possible in this framework: insertion (one state is inserted into the sequence), deletion (one state is deleted from the sequence) and substitution (one state is replaced by another state). A specific cost can be assigned to each of these elementary operations.  Typically, insertion and deletion are assigned a cost of 1 while substitution costs are allowed to vary. Using this approach, the distance between two sequences can thus be defined as the minimum cost to transform one sequence into the other. However, specifying these costs often involves subjective choices, which may lead to violations of the triangle inequality if not properly tuned. Several proposals in the literature introduced criteria to improve or guide the choice of costs in OM. For instance, estimating the substitution-cost matrix in a data-driven fashion using the between-states transition rates, i.e. using substitution costs that are inversely proportional to transition frequencies between two states \cite{Tbis}. Consider two states \textit{a} and \textit{b}. Let $N_t(a)$ and $N_t(b)$ be the number of individuals experiencing respectively \textit{a} and \textit{b} at time $t$, and $N_{t,t+1}(a,b)$ be the number of individuals experiencing \textit{a} at time $t$ and \textit{b} at time $t+1$. The transition frequency from \textit{a} to \textit{b} is \begin{equation} p_{t, t+1}(a,b) = \frac{\sum_{t=1}^{T-1}N_{t, t+1}(a,b)}{\sum_{t=1}^{T-1}N_{t}(a)}    
\end{equation}
The cost of substituting \textit{a} for \textit{b}, $c(a,b)$ can be defined as $c(a,b) = c(b,a) = 2-p_{t, t+1}(a,b)-p_{t,t+1}(b,a)$. This cost specification takes into account the occurrence of the events weighting more those transitions that are less frequent. Due to the fact that transitions at different times are qualitatively different, Lesnard (2006) proposed a dynamic hamming distance that is based on time varying substitution costs $c_t(a,b)$ \cite{Sbis}.
In addition, in settings different from bioinformatics, these edit-operations do not have a direct interpretation and this makes it difficult to obtain meaningful results. In particular, insertion and deletion operations warp time to match identically coded states but occurring at different moments in their respective sequences. A simple solution to avoid insertion and deletion operations is to use the Hamming distance that measures the minimum number of substitutions required, maintaining the original timescale \cite{T}. Although the focus on substitution operations has the sociological advantage of targeting trajectories with simultaneous similarities, in contrast to the prohibited insertion and deletion operations, which focus on matching states irrespective of their timing, this distance is liable to suffer from temporal rigidity, since anticipations and/or postponements of the same choices in life courses are not accounted for. Hence, similar sequences shifted by one time period may be maximally distant from one another, while misalignment is less of a concern for sequences exhibiting long durations in the same state.
Alternative dissimilarity criteria have also been introduced to allow control over the importance assigned to the characteristics of the sequences (namely, the collection of experienced states, their timing or their duration) in the assessment of their differences \cite{Ubis}.

\begin{figure}[ht]
    \centering
    \includegraphics[scale=0.60]{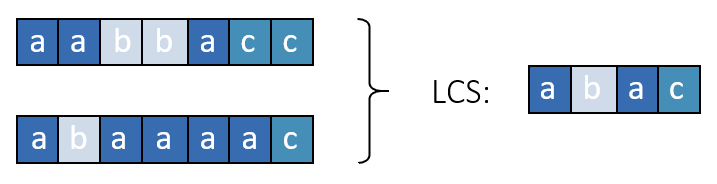}
    \caption{The Longest Common Subsequence between two sequences}
    \label{fig:2}
\end{figure}

\hspace{20pt} Later, the longest common subsequence (LCS) metric was introduced as a special case of the OM \cite{S}. According to Elzinga’s proposal, two sequences are very similar if they have in common long subsequences, while allowing gaps in between elements \ref{fig:2}. This can be use to extract stable patterns in the sequences and in this way, the length of common subsequences can be used as an indicator of the similarity of two strings. The main difference is that the OM distances identifies the part of a sequence that have to be modified to make it equal to another sequence, and LCS identifies the parts that are already sequentially common for both sequences. Therefore, the results applying OM between two sequence is a number measuring how similar two pattern are, and the results of applying LCS is a subset of the states that belong to the common subsequence extracting stable patterns followed by individuals. However, if the successive step is to cluster the sequences, we need in any case to build a matrix of distance, and the basic OM (with substitution cost equal to 2) and LCS will be the same.  Let $A(x, y)$ be a count of common attributes between sequences $x$ and $y$ that occur in the same order. It is a proximity measure since the higher the counts, the closer the sequences. We derive a dissimilarity measure from it through the following general formula: $$d(x, y) = A(x, x) + A(y, y) - 2A(x, y)$$ where $d(x, y)$ is the distance between objects $x$ and $y$. The dissimilarity is maximal when $A(x, y) = 0$; i.e., when the two sequences have no common attribute. It is zero when the sequences are identical, in which case we have $A(x, y) = A(x, x) = A(y, y)$. In particular it should be noted that a dissimilarity measure is a quantification of the distance between two sequences or, more generally, between two objects. This quantification makes it possible to compare the sequences or, for instance, to set up a dissimilarities matrix for cluster analysis purposes.

When compared with metrics based on position-wise, the LCS metric reduces distances by accounting for non-aligned matches; i.e., position-shifted similarities. Note that this methodology has a purely sequential nature and do not take into account any temporal dimensions. Even so, there are no results proving that one procedure is superior to others and the choice of dissimilarity measure remains an issue in SSE analysis. In the following, we adopted the LCS metric in order to consider similar patterns those with similar subsequences, also if translated over time.\\

After calculating a dissimilarity matrix $\textbf{D}$ obtained from a set of sequences $S = (s_1,..., s_n)$, where $n$ is the number of subjects, cluster analysis can be applied to group sequences and identify the most typical trajectories experienced by the sampled individuals. Heuristic clustering algorithms, either hierarchical or partitional, are typically used.
All cluster analyses produce results, whatever their pertinence. It is therefore necessary to discuss their quality so as to specify the scope of the results and not make unwarranted generalizations.
We conducted a cluster analysis by using a hierarchical clustering method, i.e, the Ward algorithm \cite{Z}. %Hierarchical agglomerative clustering algorithms proceed by sequentially joining pairs of clusters, and they differ in the criterion that is followed to select which clusters must be joined \cite{AA}. In Ward’s algorithm the two clusters to be joined are selected by minimizing the increase in the within-groups dispersion consequent on the reduction of the number of partitions. The optimal number of clusters can be chosen either empirically for example by visual inspection of the dendogram, or theoretically.  The theoretical methods include direct methods and statistical testing methods. Direct methods consist of optimizing a criterion, such as the within cluster sums of squares or the average silhouette. The corresponding methods are named elbow \cite{BB} and silhouette methods \cite{CC}, respectively. Statistical testing methods consist of comparing evidence against null hypothesis. An example is the gap statistic \cite{DD}. Generally several different algorithms and different numbers of groups are tested. For each of the groupings obtained, the quality of the clustering is then computed. These measures make it possible to guide the choice of a particular solution and validate it.
Our aim was to conduct a posterior interpretation of the cluster analysis results to profile the groups of pattern and, as appropriate, to compute the association between the grouping obtained and other variables of interest.

\subsection{Post-hoc analysis}
In many applications, the interest lies in relating sequences to a set of covariates and/or endpoints of interest. Within this framework, the resulting groups  may then be submitted to classical inferential methods and serve as response variables or explanatory factors for regression-like models. 
For example, one may want to know whether the sequences of young individuals differ significantly from those that are older or whether the gender has an effect on the trajectories. To that purpose it is possible to test the association between sequences and stratifying factors using a chi-squared test or logistic regression. Several covariates can be included by building for example a sequence regression tree or by multifactor discrepancy-based methods to test for differences between groups \cite{HH}. With this work we conducted a post-hoc analysis trying to characterize the clusters obtained by a traditional hierarchical procedure, but any distance-based methods may be applied. 

\hspace{20pt} In addition, results of sequences mining techniques can be integrated into classical statistical models to evaluate association between specific patterns and health-related outcomes \cite{II}. In this work clusters identified by SSA during the first year after the diagnosis of schizophrenia, and groups based on the complexity measures (high/low longitudinal entropy and turbulence) were used as independent variables to predict subsequent hospitalizations in a time-to-event analysis.\\ 
A schematic graphical representation of the overall process flow is reported in Figure 3.

\begin{figure}[!h]
    \centering
    \includegraphics[scale=0.50]{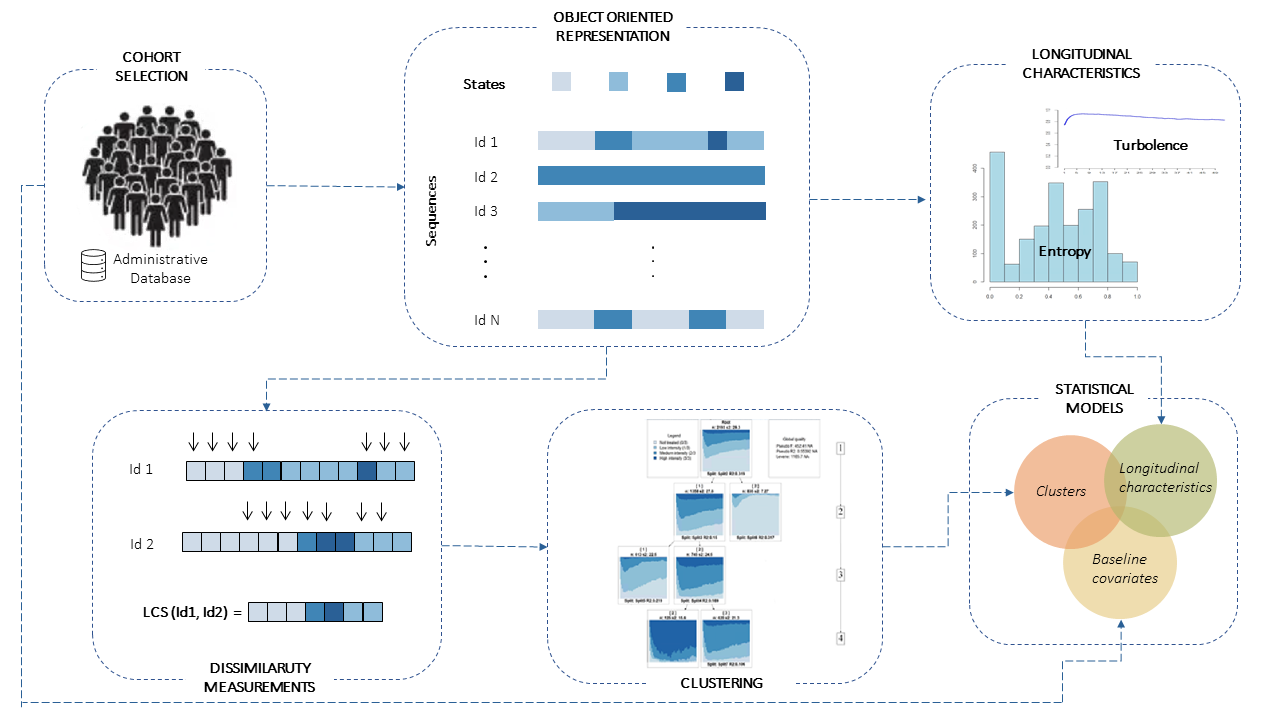}
    \caption{Algorithm pipeline of the methodology applied in this study: the SSA approach}
\end{figure}

\section{Dataset}
In this Section, we describe the administrative databases analyzed within this work (Section 4.1), and the real case-study setting (Section 4.2).

\subsection{Data sources}
The data used for this study were retrieved from the healthcare utilization databases on the residents of Lombardy, a Region of Italy that accounts for about 16\% (10 million) of its population. The Italian population is covered by the National Health Service (NHS) that provides hospitalization, major diagnostic procedures, and so-called
life-saving drugs to all citizens free or almost free of charge, including antipsychotics drugs. In addition, a specific automated system concerning psychiatric care gathers data from regional Departments of Mental Health (DMHs) accredited by the NHS. This system provides demographic information and diagnostic codes for patients receiving specialist mental healthcare \cite{EE}. Twenty-six categories of interventions and activities provided by DMHs on outpatient, home care or day care facilities were coded and classified into two broad categories: “psychosocial interventions” and “generic care”.
Because a unique identification code is used for all databases within each region, the complete care pathway of NHS beneficiaries can be obtained through record linkage. Further details on HCU database use in the field of mental health have been reported in \cite{FF, GG}.

\subsection{Study setting}
The target population of this work consisted of all NHS beneficiaries resident in Lombardy, aged 18-40 years old, who during the recruitment period (2015-2018) had at least one contact with a mental health service and had a diagnosis of schizophrenia. We then excluded prevalent patients and those with less than 1 year of follow-up. The patients included into the final cohort were followed for 1 year after the the onset of schizophrenia (i.e, without experiencing death, emigration or hospitalization during the fist year after diagnosis). During the first year we have collected all psychiatric visits, psycho-social interventions and anti-psychotic dispensed drugs delivered to each patient. The combination of these three interventions make up the optimal treatment pathway (OTP).

\section{Results}
\subsection{Definition of the states and cohort selection}

In our case-study, we defined the states of the sequence for each patient on a weekly basis. Based on the OTP, we classified each state every week into four categories, based on how many of the three treatments (i.e, psychiatric visits, psycho-social interventions and anti-psychotic) were dispensed to the patients:
\begin{itemize}
\item \textit{Not treated (0 out of 3)}, if in that given week the patient is not covered by any of
the three treatments;
\item \textit{Low variety of treatments (1 out of 3)}, if covered by only one treatment;
\item \textit{Medium variety of treatments (2 out of 3)}, if covered by two treatments out of
three;
\item \textit{High variety of treatments (3 out of 3)}, if covered by all three treatments.

\end{itemize}
In particular, we considered for each psychiatric visits a coverage of 2-4 weeks, depending on who provided the visit, for each psycho-social interventions a coverage of 2 weeks and for the anti-psychotic drugs the defined daily dose.\\

\hspace{20pt} Out of the 8'566 patients aged 18-40 years and assisted by the NHS for schizophrenia in the years 2015-2018, 2'739 were incident users. Among these, only 2'329 patients had at least 1 years of observation and were included in the final cohort. The patients' characteristics are reported in Table 1.
 In particular we reported few baseline characteristics included sex, age, and the number of co-treatment, dispensed during the two years before the date of the diagnosis of schizophrenia, was assessed and categorized as 0, 1 - 2, $\geq3$. In addition, the clinical status of the patients was quantified by the Multisource Comorbidity Score (MCS), a prognostic score that has been shown to be a good predictor of all-cause mortality and hospitalization of the Italian population \cite{IIbis}. The strata identified by the MCS are: \textit{good} (score=0), \textit{medium} (1$\leq$score$\leq$4), and \textit{poor} (score$\geq$5). Finally, we collected all the considered treatment (i.e, psychiatric visits, psycho-social interventions and anti-psychotic) during the first year of follow-up.

\begin{table}[h]
\caption{Characteristics of the cohort members}
\centering
\begin{tabular}{ll}
	\toprule
                                                                                    & \textbf{Cohort members} 
                                                                               \\
                                                      & \multicolumn{1}{c}{(N = 2,329)} \\
    \midrule
\textbf{Male   gender}                                                   & \multicolumn{1}{c}{1,449 (66.1\%)} \\
\textbf{Age class   (years at index date) }                                                  &                \\

\hspace{3pt}  18 – 29                                                           & \multicolumn{1}{c}{1,204 (54.9\%)} \\
\hspace{3pt}  30 – 40                                                           & \multicolumn{1}{c}{989 (45.1\%)}   \\

\textbf{Number of co-treatments}:                                                            &                \\
\hspace{3pt}  0                                                                  & \multicolumn{1}{c}{1,446 (65.9\%)} \\
\hspace{3pt}  1-2                                                                & \multicolumn{1}{c}{719 (32.8\%)}   \\
\hspace{3pt}  $\geq 3 $                                                            & \multicolumn{1}{c}{28 (1.3\%)}     \\
\textbf{Clinical status \small{‡}}                                                                  &                \\
\hspace{3pt}  Good                                                               & \multicolumn{1}{c}{1,065 (48.6\%)} \\
\hspace{3pt}  Intermediate                                                       & \multicolumn{1}{c}{192 (8.8\%)}    \\
\hspace{3pt}  Poor                                                               & \multicolumn{1}{c}{936 (42.7\%)}  \\
\begin{tabular}[c]{@{}l@{}}\textbf{Weekly variety of treatments \small{§}}: mean (SD)\end{tabular} &                \\
\hspace{3pt}  No treatments                                                  & \multicolumn{1}{c}{24.1 (20.2)}    \\
\hspace{3pt}  Low                                                              & \multicolumn{1}{c}{13.1 (12.6)}   \\
\hspace{3pt}   Medium                                                           & \multicolumn{1}{c}{10.9 (12.9)}    \\
\hspace{3pt}  High                                                              & \multicolumn{1}{c}{3.9 (9.6)}      \\
	\bottomrule
\multicolumn{2}{c}{\small{\small{‡} Clinical status was assessed by the MCS, and 3 categories were considered: good (score=0), }}\\
\multicolumn{2}{c}{\small{medium (1$\leq$ score $\leq$4), poor (score $\geq5$) ;\hspace{170pt} }}\\
\multicolumn{1}{c}{\small{\small{§} Treatments are considered during the first year of follow-up.}}
\end{tabular}
\end{table}
The majority of these patients were male (66.1\%), more frequently under 30 years old. We also notice a low number of co-treatments, also being patients very young, but 42.7\% of them resulted to have a poor clinical status.
During the first year, on average, the patients resulted not covered by any kind if treatment for 24 weeks (almost 6 months), for about a month they were covered by 1 or 2 kind of treatment, and only for a month the variety of treatments resulted to be high, meaning that the patients are fully covered with all the OTP's treatments. 

\subsection{Sequences' characterization}
The transition rates between states, as defined in Section 5.1, are showed in Table 2. The most stable state is the \textit{no treatment} one, (\textit{0 out of 3}), with a probability of remaining in this state, for those who are already there, equal to 95\%. On the other hand, for the other 3 states, the probability of remaining in the same state is 81\%, 83\% and 87\%, respectively. 
Furthermore, we see the highest transition rate for the patient that discontinued a treatment, i.e., from \textit{2 out of 3} to \textit{1 out of 3} and from \textit{3 out of 3} to \textit{2 out of 3} state, that is equal to 12\%. Finally, as expected, patients tend to add or discontinue a treatment at once, tending the transition rate from \textit{3 out of 3} to \textit{0 out of 3} or \textit{1 out of 3} state to zero.\\

\begin{table}[h]
\centering
\caption{Transition Matrix}
\label{tab:my-table}
\begin{tabular}{ccccc}
	\toprule
                        & -\textgreater   (0/3) & -\textgreater   (1/3) & -\textgreater   (2/3) & -\textgreater   (3/3) \\
                 	\midrule
(0/3)   -\textgreater{} & 0.95                  & 0.04                  & 0.01                  & 0.00                  \\
(1/3)   -\textgreater{} & 0.09                  & 0.81                  & 0.10                  & 0.01                  \\
(2/3)   -\textgreater{} & 0.01                  & 0.12                  & 0.83                  & 0.04                  \\
(3/3)   -\textgreater{} & 0.00                  & 0.01                  & 0.12                  & 0.87                  \\
	\bottomrule
\end{tabular}
\end{table}

\begin{figure}[ht]
    \centering
    \includegraphics[scale=0.50]{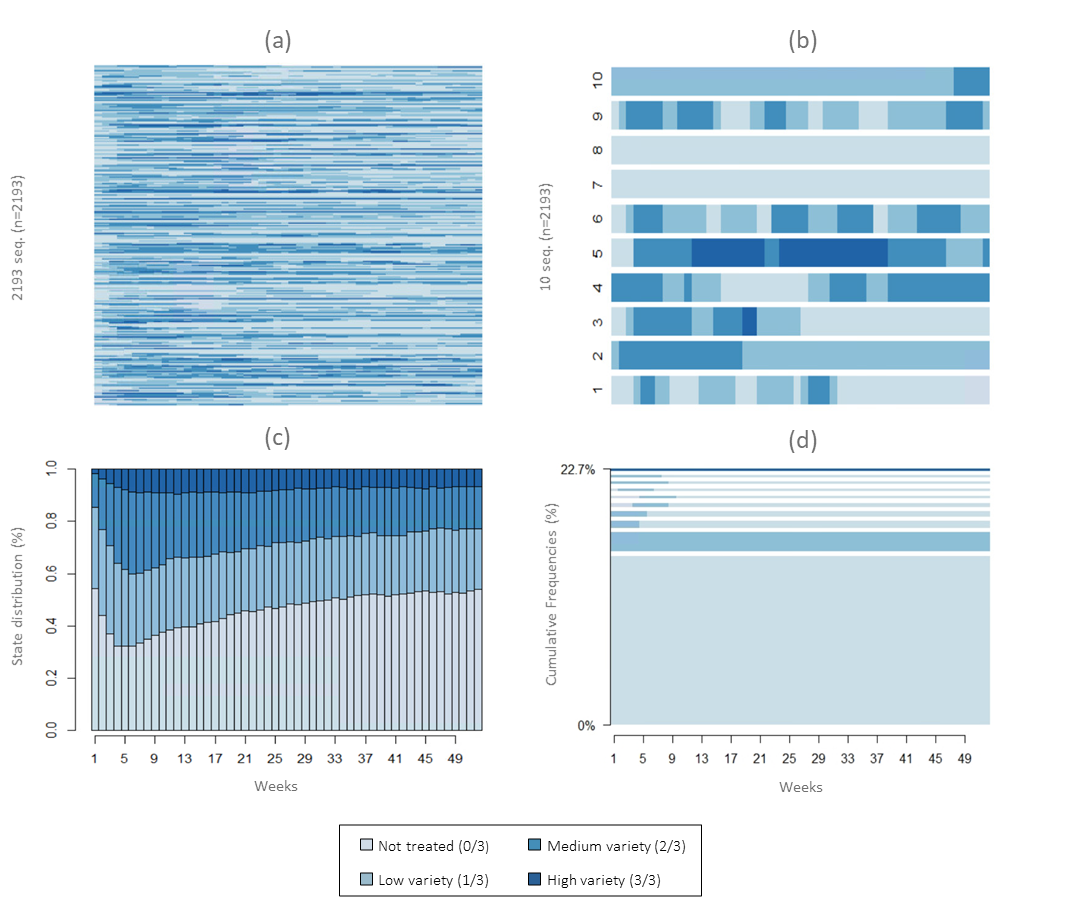}
    \caption{Different visualization of sequences: (a) Index plot; (b) Sequences of the first ten patients; (c) State distribution plot; (d) Frequency plot.}
    \label{fig:3}
\end{figure}
In the top panel of Figure 4, the sequences of each cohort member (a) and of the first 10 patients (b) are displayed. The high variability between patients and the inadequacy of the most of the pattern can be noticed. The state distribution plot (c) shows a high proportion of patients without any treatments during all the weeks, being almost always equal to or greater than 50\%. Only around the end of the first month the proportion of patient treated with \textit{2 out 3} treatments seems to be increased, but soon many patients discontinue the treatments initiated. During the first year after the diagnosis of schizophrenia, 18\% of patients did not received any treatment, and with the exception of the few patients who remain treated with at least one treatment throughout the year, plot (d) shows many sequences representing patients who started a treatment around the first month of follow-up, but then have it soon discontinued.

\subsection{Profiling sequences}
The calculation of the distance through the LCS metric and the study of the optimal number of groups to be considered allowed us to cluster our patients into 3 different groups.

Table 3 shows the distributions of variables across clusters, that can be used to perform an a-posteriori profiling of the clusters. Male and female are equally distributed among clusters, whereas the third cluster resulted to be more represented by young patients, but with worst clinical conditions and with the higher percentage of co-treatments, despite their young age. Finally, we can characterize the 3 groups as:

\begin{itemize}
    \item 1,237 patients with \textit{very low variety} of treatment [1];
    \item 662 patients with \textit{medium-low variety} of treatment [2];
    \item 294 patients with \textit{medium-high variety} of treatment [3].
\end{itemize}

\begin{table}[h]
\centering
\caption{Characteristics of the cohort members by clusters}
\begin{tabular}{lcccc}
	\toprule
                                                                                     & \textbf{Cluster 1} & \textbf{Cluster 2} & \textbf{Cluster 3} & \textbf{P-value} \\
   & (N = 1,237) & (N = 662) & (N = 294)       \\                                                                            
\midrule
\textbf{Male gender}                                                                 & 799 (64.6\%)       & 443 (66.9\%)       & 207 (70.4\%)       & 0.143            \\
\textbf{Age class (years at index date)}                                             &                    &                    &                    &                  \\
\hspace{3pt} 18 – 29                                                                              & 632 (51.1\%)       & 351 (53.0\%)       & 221 (75.2\%)       & \textless 0.001  \\
\hspace{3pt} 30 – 40                                                                              & 605 (48.9\%)       & 311 (47.0\%)       & 73 (24.8\%)        &                  \\

\textbf{Number of co-treatments:}                                                    &                    &                    &                    &                  \\
\hspace{3pt} 0                                                                                    & 866 (70.0\%)       & 407 (61.5\%)       & 173 (58.8\%)       & \textless{}0.001 \\
\hspace{3pt} 1-2                                                                                  & 351 (28.4\%)       & 247 (37.3\%)       & 121 (41.2\%)       &                  \\
\hspace{3pt} $\geq 3$                                                                                  & 20 (1.6\%)         & 8 (1.2\%)          & 0 (0.0\%)          &                  \\
\textbf{Clinical status \small{‡} }                                                            &                    &                    &                    &                  \\
\hspace{3pt} Good                                                                                 & 636 (51.4\%)       & 324 (48.9\%)       & 105 (35.7\%)       & \textless{}0.001 \\
\hspace{3pt} Intermediate                                                                         & 121 (9.8\%)        & 57 (8.6\%)         & 14 (4.8\%)         &                  \\
\hspace{3pt} Poor                                                                                 & 480 (38.8\%)       & 281 (42.5\%)       & 175 (59.5\%)       &                 \\
\begin{tabular}[c]{@{}l@{}} \textbf{Weekly variety of treatments \small{§}}:  mean (SD)\end{tabular}  &                    &                    &                    &                  \\
\hspace{3pt} No treatments                                                                        & 39.6 (12.2)        & 5.0 (5.9)          & 1.9 (3.0)          & \textless{}0.001 \\
\hspace{3pt} Low                                                                                  & 9.0 (9.6)          & 23.0 (13.7)        & 7.5 (6.5)          &                  \\
\hspace{3pt} Medium                                                                               & 2.9 (5.0)          & 22.8 (13.9)        & 17.9 (8.8)         &                  \\
\hspace{3pt} High                                                                                 & 0.5 (2.2)          & 1.2 (2.9)          & 24.7 (12.2)        &                  \\
\bottomrule
\multicolumn{5}{c}{\small{\small{‡} Clinical status was assessed by the MCS, and 3 categories were considered: good (score=0), medium (score 1 - 4),}}\\
\multicolumn{1}{c}{\small{ poor (score$\geq5$); \hspace{155pt}}} \\
\multicolumn{2}{c}{\small{\small{§}Treatments are considered during the first year of follow-up  \hspace{20pt} }}
\end{tabular}
\end{table}

\begin{figure} [h]
\centering
    \includegraphics[scale=0.45]{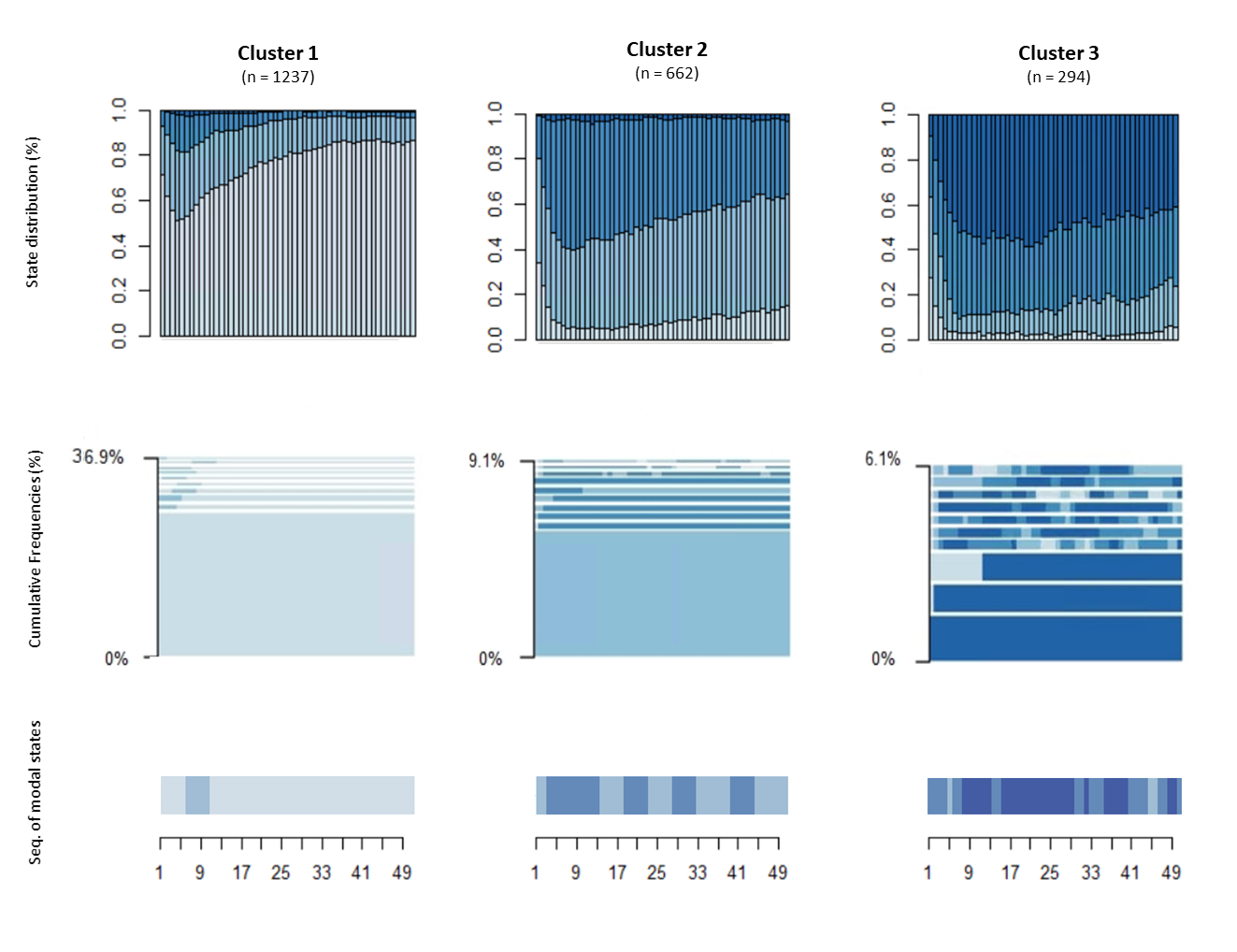}
    \caption{ State distribution plots (first row), frequency plots (second row), and Sequence of modal states plots (third row) by clusters.}
    \label{fig:3}
\end{figure}

The state distribution plots, the frequency plot and the sequence made of the modal state at each position, i.e., the typical trajectory, for each cluster (Figure 5) show the clear differences between patterns followed by the patients of different groups.

\subsection{Outcome measure}
\hspace{20pt} The results of the cluster analysis were then embraced to predict the time of relapse occurrences, such as  an emergency hospital admissions reporting a diagnosis of schizophrenia spectrum disorder or an admission in a psychiatric ward. 
They were labelled as outcome episodes and were considered as measurable surrogates of relapse \cite{IIbis2, IIbis3}, although the question is still open and debated.
The results of the fitted Cox model are reported in Table 4. Compared to patients with a very low variety of treatment (cluster 1), those with medium-low and medium-high variety of treatments showed an increment in the adjusted risk of hospitalization of 80\% (95\%CI 43\%-126\% ) and 62\% (95\%CI 21\%-118\%), respectively. Being the longitudinal entropy and the turbulence widely explained already by clusters, the probability of being hospitalized seems to be not influenced by them (see the adjusted model). On the contrary, in the first column where the univariable HRs are reported we see both the clusters and the indices being significantly associated with the outcome.

\begin{table}[h!]
\centering
\caption{Univariable and adjusted hazard ratio (HR), and 95\% confidence intervals (CI), of first hospitalization}
\begin{tabular}{lcc}
	\toprule
\multicolumn{1}{c}{} & Univariable HR (95\%CI)   & Adjusted HR (95\%CI) \\
\midrule
Cluster 2 \small{‡}          & 1.81 (1.47 - 2.23) & 1.80 (1.43 - 2.26)  \\
Cluster 3 \small{‡}            & 1.93 (1.49 - 2.51) & 1.62 (1.21 - 2.18)  \\
High Entropy \small{§}         & 1.36 (1.13 - 1.64) & 1.07 (0.83 - 1.38)  \\
High Turbulence \small{§}      & 1.37 (1.13 - 1.66) & 0.93 (0.70 - 1.21) \\
\bottomrule
\multicolumn{3}{c}{\small{\small{‡} Cluster 1 is the reference; \small{§} Low entropy and low turbulence i.e below the mean, are the references } }
\end{tabular}

\end{table}

\newpage
\section{Discussion and Conclusions}
The aim of the present study was to provide a methodological pipeline based on state sequences for the analysis and profiling of longitudinal care pathways. This methods allows to describe the entire pathway of care of each patients, as statistical object, and provide both visual and numerical instruments for their evaluation and connection with the main endpoint of interest.  Furthermore, SSA does not make any assumptions about the data-generating mechanism, allowing a more proper description and mining of the events.

\hspace{20pt} In pharmacoepidemiology practice it is common that predictive models use fixed baseline measures for the consumption of the health care \cite{IIbisbis, IIbisbisbis}. However, these indices, such as the adherence to drug therapy or to guidelines' recommendations, discard valuable information. They indeed do not take into account the whole pattern evolution of each patient, being not informative as it may be. For these reasons, SSA is a way to properly consider all the longitudinal processes that characterize a single patients as a mathematical object featuring the single statistical unit, with the aim of summarize this complex information with some features that can be inserted into standards predictive models and are able to reflect the dynamics and the behavior of the patients. This kind of approach results more realistic and informative with respect to the commonly used baseline measures, as was the attempt of some authors that introduced the dynamic monitoring of the effects of adherence to medication effect \cite{LL}.

\hspace{20pt} Our working case can be a valid example to how extrapolate useful and easy to read information of the patients' pathway. It can be concluded that the method adopted allows to quantify the phenomenon longitudinally leading to a quantitative identification of the assistance gap still present for these patient. This implies that data can really seen as a feed-back tool of real clinical practice for decision-making processes. However, although evidence is needed on effective treatment sequences, the main limitation actually present in the field selected for the application is that it is hard to define a real indicator of efficacy of the therapeutic pathway. This is due to the difficulty of finding a shared definition of the adverse outcome that can really be indicative of a worsening of mental illness. It is indeed necessary to validate the clusters obtained, i.e., to verify whether an intensification of treatments results in better outcomes, and to identify the appropriate outcomes for this evaluation.
The clusters that have been identified seem to really group patients with more similar pathways, but in a counter-intuitive way we obtained that the patients more treated showed an higher risk of hospitalization. Actually, considering the impossibility of adjusting the Cox regression for relevant data such as the severity of schizophrenia at cohort entry and lifestyle factors (among others), it is likely that a cohort approach could generate estimates affected by confounding. In other words, as pointed out by Corrao et al \cite{MM}, because patients who have more severe schizophrenia at baseline are more likely to receive timely and continuative care, but also more frequently experience relapse, a paradox positive association between intensity of mental healthcare and outcome occurrence may be generated from this simple design. However, the SSA method can be easily applied within the traditional cohort designs and case-control studies, as well as the self-matched designs such as case-crossover or self controlled case series \cite{NN, OO,PP, QQ}. From the mental health point of view, this work helped to stress some important concept: (i) data-driven techniques may potentially support the empirical identification of effective care sequences by extracting them from data collected regularly in healthcare, (ii) SSA can be a tool capable of providing insights on the interval and timing of treatment patterns used in practice and on the effectiveness of different treatment sequences. This knowledge could be used for monitoring, evaluating and possibly (re)designing optimizing existent care pathways. However, it is necessary to identify adequate indicators having the availability of qualitatively acceptable data able to implement them. 

\hspace{20pt} This research represents a first important step in the assessment of care pathways and paves the way for many further developments. Among others, model-based approach like latent-class analysis \cite{QQbis}, which enables a direct modelling of the sequences (intended as objects) \cite{QQbis2}, and finite mixture of exponential-distance model \cite{QQbisbis}, which allows covariates to affect the soft cluster membership probabilities, rather than using them for the a-posteriori profiling of the clusters. This second approach is highly of interest when the main objective is to better understand if and to what extent the typical sequence patterns characterising each cluster are affected by specific covariates.
Moreover, a design study to properly validate the clusters has to be found. Furthermore, SSA is a flexible methodology and thanks to the multiple options to define states, time granularity and metrics used for the distances it can be adapted to different epidemiological contexts. In addition, an interesting development in sequence analysis is the so called multichannel sequence analysis; this method is an extension of the conventional SSA presented in this article, and it allows to describe individual trajectories on several dimensions simultaneously \cite{RR}.

\hspace{20pt} In conclusion, the analysis of state sequences is useful and effective in pharmacoepidemiologic studies because it offers a valuable gain in patterns discovery. Furthermore, via the inclusion of covariates, it could establish why certain sequence patterns exist and allow the researcher to capture the potential predictive effect that a specific treatment pathway has on future outcomes. Therefore, SSA could be an effective tool supporting providers to assess the quality and effectiveness of the diagnostic-therapeutic pathways provided to these patients and to monitor them over time. 

\newpage
%%%%%%%%%%%%%%%%%%%%%%%% referenc.tex %%%%%%%%%%%%%%%%%%%%%%%%%%%%%%
% sample references
% %
% Use this file as a template for your own input.
%
%%%%%%%%%%%%%%%%%%%%%%%% Springer-Verlag %%%%%%%%%%%%%%%%%%%%%%%%%%
%
% BibTeX users please use
% \bibliographystyle{}
% \bibliography{}
%

\end{document}